\documentclass[%
 reprint,
groupedaddress,
nobibnotes,
amsmath,amssymb,
aps,
prl,
]{revtex4-2}

\usepackage{graphicx}
\usepackage{dcolumn}
\usepackage{bm}
\usepackage{color}
\usepackage{braket}
\usepackage[normalem]{ulem}
\usepackage{comment}

\begin{document}

\preprint{APS/123-QED}

\title{Coherence squeezing in optical interference}

\author{Martti~Hanhisalo}
\email{martti.hanhisalo@uef.fi}
\affiliation{%
 Center for Photonics Sciences, University of Eastern Finland, P.O. Box 111, FI-80101 Joensuu, Finland
}%
\author{Atri~Halder}
\affiliation{%
 Center for Photonics Sciences, University of Eastern Finland, P.O. Box 111, FI-80101 Joensuu, Finland
}%
\author{Tero~Set\"{a}l\"{a}}%
\affiliation{%
 Center for Photonics Sciences, University of Eastern Finland, P.O. Box 111, FI-80101 Joensuu, Finland
}%
\author{Andreas~Norrman}
\affiliation{%
 Center for Photonics Sciences, University of Eastern Finland, P.O. Box 111, FI-80101 Joensuu, Finland
}%

\date{\today}

\begin{abstract} 
We introduce the concept of optical coherence squeezing in double-slit interference. We construct Hermitian operators that characterize the coherence at the slits, leading to coherence uncertainty relations and a corresponding squeezing condition. We also analyze states exhibiting such squeezing and show its manifestations in the uncertainty of the magnitudes and positions of the intensity fringes. Our work identifies coherence as a fundamental degree of freedom for squeezing, complementing phase, amplitude, and polarization, which could benefit quantum-enhanced interferometry.
\end{abstract}

\maketitle


\emph{Introduction.}---Interference and coherence are among the most fundamental and tightly linked notions in classical and quantum wave physics. In optics, the double-slit experiment reveals how the coherence of light dictates the formation of interference fringes~\cite{BornWolf}. Coherence also governs diffraction, propagation, and light-matter interactions~\cite{Mandel:95}, and serves as a key resource in imaging, spectroscopy, and photonic information processing~\cite{Whitehead:09,Clark:12,De:21,Bourassin-Bouchet:15,Dong:24}. In quantum optics, the theory of optical coherence~\cite{Glauber:63} underpins interferometry, metrology, and quantum information protocols~\cite{Pleinert:21,Su:17,Fortsch:13,Deng:19}, and its concepts appear across diverse platforms such as cosmology, quantum electronics, optomechanics, and molecular physics~\cite{Giovannini:11,Brange:15,Cohen:15,Rosenberg:22}.
 
Quantum uncertainty and squeezing are central to high-precision optical measurements. Quantum noise in amplitude, phase, and polarization of light~\cite{Agarwal:13} limit sensitivities in metrology and polarimetry~\cite{Giovannetti:06,Luis:16,Soto:21,Goldberg:22}, while appropriately engineered squeezed states can surpass these limits. Phase-squeezed light, for example, has become indispensable in quantum-enhanced gravitational-wave detection~\cite{Aasi:13,Tse:19}, with further benefits in biological sensing, spectroscopy, and dark-matter searches~\cite{Bowen:13,Taylor:16,Belsley:23,Herman:25,Backes:21}. Despite this broad relevance, the quantum uncertainty of optical coherence has only recently been explored~\cite{Hanhisalo:24}. Given the vital role of coherence in interference, a natural question arises: How do quantum fluctuations of coherence constrain---and potentially enhance---optical interferometry?

In this Letter, we introduce the concept of coherence squeezing in optical two-slit interference. We construct operators that characterize the statistical properties of light in the slit and observation planes. These operators lead to coherence uncertainty relations and a formal basis for defining coherence squeezing. We then study bright light and single-photon light that exhibit such squeezing and represent it via a geometrical construction akin to the Poincar\'e sphere. Finally, we show how coherence squeezing affects the uncertainties of the magnitudes and positions of the intensity fringes. Our work thus suggests an uncharted quantum degree of freedom for squeezing that could find use in high-precision interferometry.

\emph{Squeezing of optical coherence.}---Let two illuminated slits (pinholes) be located in an opaque screen $\mathcal{A}$. The emerging light of angular frequency $\omega$ is observed on a screen $\mathcal{B}$ at time $t$. Under typical conditions~\cite{Gerry}, the positive frequency part of the electric field operator at $\mathcal{B}$ can be written as $\hat{E}=K(\hat{a}_1\mathrm{e}^{\mathrm{i}kr_1}+\hat{a}_2\mathrm{e}^{\mathrm{i}kr_2})\mathrm{e}^{-\mathrm{i}\omega t}$. Here $K$ is a constant, $k$ is the wave number, $\hat{a}_m$ is the annihilation operator at slit $m\in\{1,2\}$, and $r_m$ is the slit distance to the observation point. The first-order statistical properties of the fields at $\mathcal{A}$ can be characterized by four Hermitian operators:
\begin{equation}
\begin{gathered}\label{G-operators-monochromatic}
   \hat{G}_0=\hat{a}_1^\dagger\hat{a}_1+\hat{a}_2^\dagger\hat{a}_2,\quad\hat{G}_1=\hat{a}_1^\dagger\hat{a}_1-\hat{a}_2^\dagger\hat{a}_2,\\
    \,\,\,\,\,\,\hat{G}_2=\hat{a}_1^\dagger\hat{a}_2+\hat{a}_2^\dagger\hat{a}_1,\quad\hat{G}_3=\mathrm{i}(\hat{a}_2^\dagger\hat{a}_1-\hat{a}_1^\dagger\hat{a}_2).
\end{gathered}
\end{equation}
We call $\hat{G}_n$, with $n\in\{0,1,2,3\}$, the G operators, as their expectation values $G_n=\langle\hat{G}_n\rangle$ correspond to the classical G parameters~\cite{Halder:21}. The parameter $G_0$ represents the total intensity of the two fields, whereas $G_1$ describes their intensity difference. Moreover, together $G_2$ and $G_3$ express the mutual coherence of the fields, being connected to the usual coherence function via $\Gamma_{12}= G_2/2+\mathrm{i}G_3/2$~\cite{Mandel:95}.

The G operators share the mathematical structure of the Stokes operators~\cite{Soto:21, Luis:16} and satisfy the commutators
\begin{subequations}
\begin{equation}
    \!\!\![\hat{G}_0,\hat{G}_j]=0,\enspace\, [\hat{G}_j,\hat{G}_k]=2\mathrm{i}\epsilon_{jkl}\hat{G}_l,\label{commutators}
\end{equation}
where $\epsilon_{jkl}$ is the Levi--Civita tensor with $ j,k,l\in\{1,2,3\}$. The latter relation in Eq.~(\ref{commutators}) precludes the exact simultaneous knowledge of $G_1$, $G_2$, and $G_3$, i.e., the information on the intensity difference and coherence. 
In particular, we obtain the uncertainty relations
\begin{equation}
    \Delta G_j\Delta G_k\geq|\epsilon_{jkl}||G_l|,\label{uncertainty-relations}
\end{equation} 
\end{subequations}
where $\Delta G_{j}=(\braket{\hat{G}_{j}^2}-\langle\hat{G}_{j}\rangle^2)^{1/2}$ is the standard deviation of $G_j$. One relation governs solely coherence uncertainties:
\begin{equation}
    \Delta G_2\Delta G_3\geq |G_1|.\label{first uncertainty relation}
\end{equation}
Accordingly, whenever the light intensities at the slits differ ($G_1\neq 0$), $G_2$ and $G_3$ must show uncertainty, viz., the coherence properties cannot be assessed with arbitrary precision. The two other relations in Eq.~(\ref{uncertainty-relations}) are
\begin{equation}
    \Delta G_1\Delta G_2\geq|G_3|,\quad \Delta G_1\Delta G_3\geq|G_2|.\label{dual uncertainty relations}
\end{equation}
Together, they state the following: If any correlations exist between the slit fields ($G_2\neq0$, or $G_3\neq0$, or both), uncertainty necessarily occurs in the intensity difference ($G_1)$ and the coherence properties ($G_3$ or $G_2$, or both). 

The uncertainty relations in Eqs.~(\ref{first uncertainty relation}) and (\ref{dual uncertainty relations}) set fundamental precision limits on the characterization of coherence and intensity difference at the slits. Moreover, they lead to the nonclassical concept of coherence squeezing, which we will discuss next.

We select the two-mode coherent state $\ket{\alpha}_1\!\ket{\beta}_2$ as the reference point for coherence squeezing. For this state, the standard deviations of the G operators in Eq.~(\ref{G-operators-monochromatic}) are all equal: $\Delta G_j=\sqrt{|\alpha|^2+|\beta|^2}=\sqrt{\bar{n}}$ with $j\in\{0,1,2,3\}$. Here, $\bar{n}$ is the total average photon number. 
When the light exhibits fluctuations in $G_2$ or $G_3$ less than this Poissonian-type noise,
\begin{equation}
    \Delta G_2 <\sqrt{\bar{n}}\quad\mathrm{or}\quad \Delta G_3<\sqrt{\bar{n}},\label{squeeze-criterion}
\end{equation}
we regard it as coherence squeezed. Next, to elaborate how such squeezing is linked to the intensity fringes, we formulate an interference law via the G operators.

We employ Eq.~(\ref{G-operators-monochromatic}) and express the intensity operator $\hat{I}=\hat{E}^\dagger\hat{E}$ at screen $\mathcal{B}$ as
\begin{subequations}
\begin{equation}
    \hat{I}=|K|^2[\hat{G}_0+\hat{G}_2\cos(k\Delta r)-\hat{G}_3\sin(k\Delta r)].\label{interference-law}
\end{equation}
Here, $\Delta r=r_2-r_1$ is the path difference. The mean of Eq.~(\ref{interference-law}) coincides with the classical interference law~\cite{Mandel:95}:
\begin{align}
    I=|K|^2\big\{G_0&+|G_2+\mathrm{i}G_3|\nonumber\\
    &\times\cos[\arg(G_2+\mathrm{i}G_3)+k\Delta r]\big\}.\label{B-intensity}
\end{align}
\end{subequations}
It demonstrates how the intensity at $\mathcal{B}$ depends both  on the total intensity at the slits ($G_0$) and the interference term ($G_2$ and $G_3$). The magnitude $|G_2+\mathrm{i}G_3|$ determines the strength of the intensity modulation, while the phase $\arg(G_2+\mathrm{i}G_3)$ specifies the fringe locations. When normalized with the total intensity, the magnitude yields the visibility of the fringes: 
\begin{subequations}
\begin{equation}
   V=\frac{\sqrt{G_2^2+G_3^2}}{G_0}.\label{visibility}
\end{equation}
The parameter $G_1$, in turn, is linked to the intensity distinguishability~\cite{Norrman:17,Norrman:20} of the light at the slits via
\begin{equation}
    D=\frac{|G_1|}{G_0}.\label{distinguishability}
\end{equation}
\end{subequations}
Since the relation $(G_1^2+G_2^2+G_3^2)^{1/2}\leq G_0$ holds~\cite{Halder:21}, the identity $V^2+D^2\leq1$  readily follows from Eqs.~(\ref{visibility}) and (\ref{distinguishability}).

\begin{figure}
    \centering
    \includegraphics[width=0.85\linewidth]{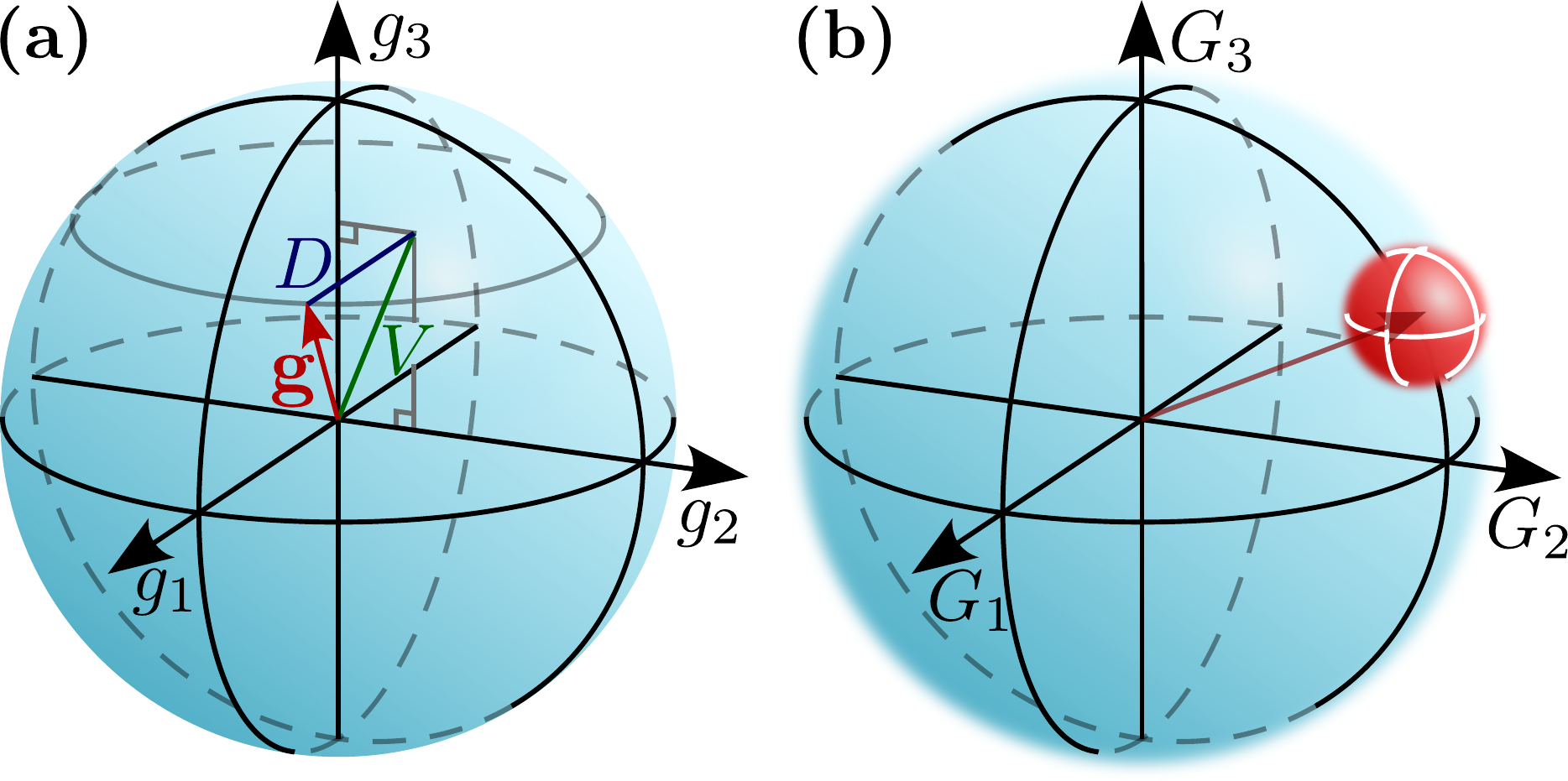}
    \caption{Illustration of the G space. \textbf{(a)} Normalized interference Poincar\'e sphere with the state vector $\mathbf{g}$, intensity distinguishability $D$, and fringe visibility $V$. \textbf{(b)} Representation of a two-mode coherent state $\ket{\alpha}_1\!\ket{\beta}_2$. The sphere has an average radius of $G_0$. The uncertainties of the G parameters are equal, as depicted by the uniform uncertainty cloud (red).}
    \label{Fig1}
\end{figure}

Assigned to three orthogonal coordinate axes, $G_1$, $G_2$, and $G_3$ span the so-called G space~\cite{Halder:21}. In this space, the vector $\mathbf{G}=(G_1,G_2,G_3)$ is located within the interference Poincar\'e sphere; The tip of the intensity-normalized vector $\mathbf{g}=\mathbf{G}/G_0=(g_1,g_2,g_3)$ lies within a unit sphere. Figure~\ref{Fig1}(a) illustrates such a configuration where the length of the $\mathbf{g}$ vector's projection on the $g_1$ axis is $D$ and on the $g_2g_3$ plane is $V$. In Fig.~\ref{Fig1}(b), we visualize the uniform G parameter uncertainties of the coherent state via the unnormalized sphere.

Before considering specific light states that exhibit coherence squeezing, we rewrite the uncertainty relations in Eqs.~(\ref{first uncertainty relation}) and (\ref{dual uncertainty relations}) in terms of $D$ and $V$. The former, by using Eq.~(\ref{distinguishability}), takes the form
\begin{subequations}
\begin{equation}
    \frac{\Delta G_2\Delta G_3}{G_0}\geq D.\label{coherence-uncertainty-relation}
\end{equation}
The intensity distinguishability thus limits the characterization precision of coherence ($G_2$ and $G_3$). The visibility $V$ in Eq.~(\ref{visibility}) sets the lower bound for another inequality, obtained by combining the relations in Eq.~(\ref{dual uncertainty relations}):

\begin{equation}
     \frac{\Delta G_1\sqrt{(\Delta G_2)^2+(\Delta G_3)^2}}{G_0}\geq V.\label{visibility uncertainty relation}
\end{equation}   
\end{subequations}
From a measurement perspective, a non-zero fringe visibility implies the presence of fluctuations of $G_1$ together with $G_2$ and/or $G_3$ at the slits.

\emph{Bright squeezed light.}---Consider a two-mode state composed of similar but independent displaced squeezed states $\ket{\Psi}=\hat{D}_1(\alpha)\hat{D}_2(\alpha)\hat{S}_1(\xi)\hat{S}_2(\xi)\ket{0}_1\!\ket{0}_2$~\cite{Korolkova:02}. Here, $\hat{D}_m(\alpha)$ is the displacement operator of mode $m\in\{1,2\}$, while $\hat{S}_m(\xi)=\exp[(\xi^*\hat{a}_m^2-\xi\hat{a}_m^{\dagger 2})/2]$ is the mode's squeeze operator with $\xi=q\exp(\mathrm{i}\theta)$. The squeeze parameter $q$ and angle $\theta$ determine how the field quadratures $\hat{X}_m=(\hat{a}_m^\dagger+\hat{a}_m)/2$ and $\hat{Y}_m=\mathrm{i}(\hat{a}^\dagger_m-\hat{a}_m)/2$ are squeezed. 

As an example, let $\alpha$ be real and $\theta=0$. The expectation values of the quadrature operators are $X_m=\alpha$ and $Y_m=0$, while their uncertainties are $\Delta X_m=\mathrm{e}^{-q}/2$ and $\Delta Y_m=\mathrm{e}^{q}/2$. Clearly, the $X$ quadratures are squeezed and the $Y$ quadratures are antisqueezed. The squeezing directions align with the displacements; Hence, the light is amplitude squeezed in both slits. Such squeezing affects also the coherence properties. Assuming $\bar{n}\gg\sinh q$ and omitting some negligible terms, Eq.~(\ref{G-operators-monochromatic}) yields the mean values $G_0\approx G_2=\bar{n}$ and $G_1=G_3=0$ as well as the standard deviations
\begin{subequations}
\begin{align}
    \Delta G_0=\Delta G_1=\Delta G_2\approx\sqrt{\bar{n}}\mathrm{e}^{-q},\quad\Delta G_3=\sqrt{\bar{n}}\mathrm{e}^{q}.\label{G uncertainties}
\end{align}
We observe squeezing in $G_0$, $G_1$, and $G_2$. Squeezing in $G_0$ indicates a sharper total photon number, while for $G_1$ it implies a more precise photon number difference between the slits. More importantly, the uncertainty of $G_2$ is reduced, while the uncertainty of $G_3$ is increased. The criterion in Eq.~(\ref{squeeze-criterion}) is thus met, viz., the light is coherence squeezed in $G_2$. Figure~\ref{Fig2}(a) illustrates the squeezing in the G space and in the coherence ($G_2G_3$) plane. The latter presentation shares similarity with the usual quadrature phase-space picture~\cite{Gerry}. Here, the squeezing occurs in the magnitude $|G_2+\mathrm{i}G_3|$ while the antisqueezing takes place in the phase angle $\arg(G_2+\mathrm{i}G_3)$.

\begin{figure}
    \centering
     \includegraphics[width=0.84\linewidth]{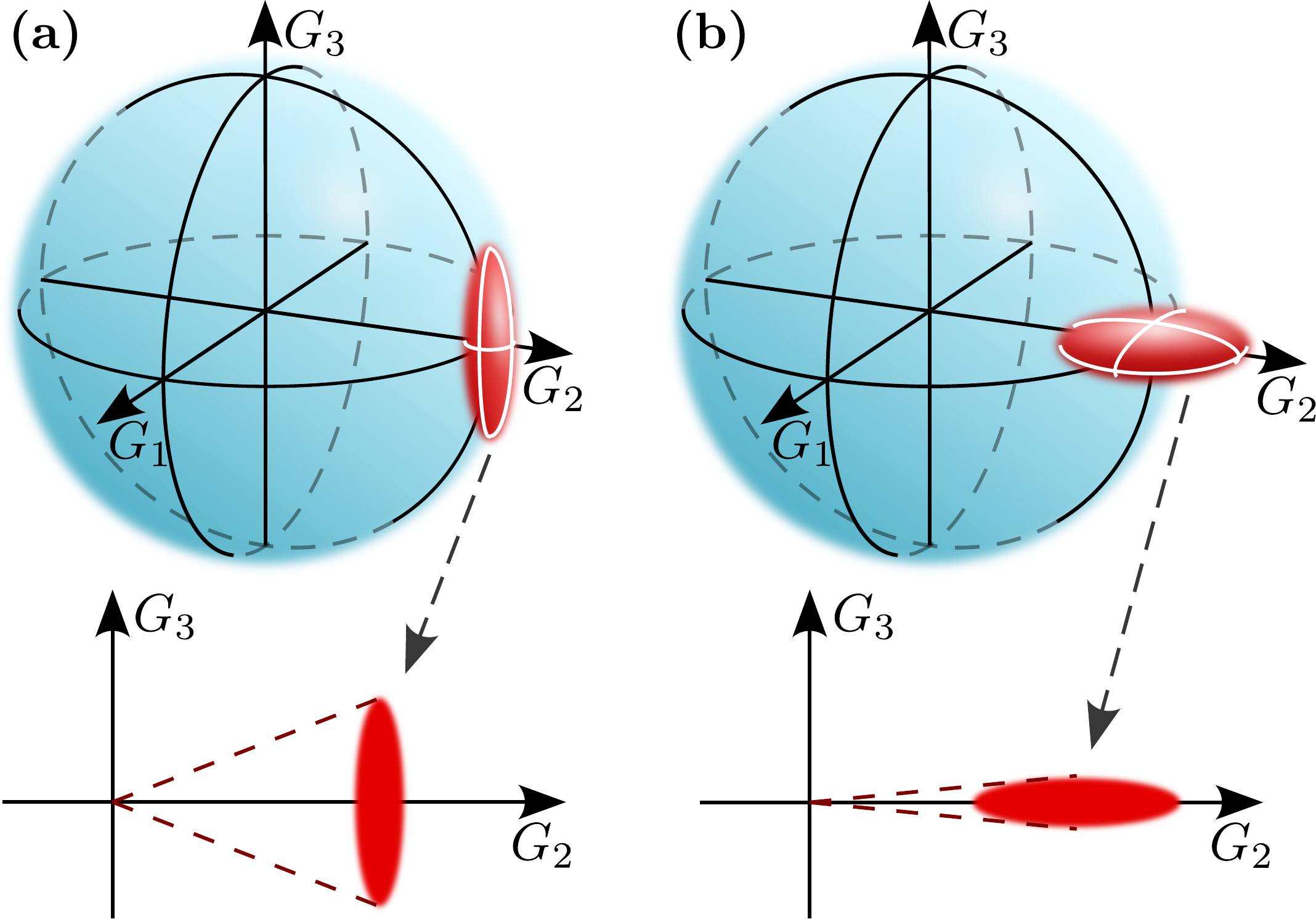}
    \caption{
    G-space representations of bright quadrature-squeezed states. In both cases, a projection of the uncertainty volume is presented in the $G_2G_3$ plane. In \textbf{(a)} the magnitude $|G_2+\mathrm{i}G_3|$ is squeezed, while in \textbf{(b)} the squeezing occurs in the phase $\arg(G_2+\mathrm{i}G_3)$.}
    \label{Fig2}
\end{figure}

Another example covers phase-squeezed light. If $\theta=\pi$, the $Y$ quadratures show squeezing: $\Delta Y_m=\mathrm{e}^{-q}/2$ and $\Delta X_m=\mathrm{e}^{q}/2$. The majority of the fluctuations now occur in the displacement directions $(X_m)$, rendering the light antisqueezed in amplitude and squeezed in phase at the slits. In this case, the G-operator expectations are the same as above Eq.~(\ref{G uncertainties}), but their fluctuations are
\begin{equation}
    \Delta G_0=\Delta G_1=\Delta G_2\approx\sqrt{\bar{n}}\mathrm{e}^{q},\quad\Delta G_3=\sqrt{\bar{n}}\mathrm{e}^{-q}.
\end{equation} 
\end{subequations}
We see that solely $G_3$ is squeezed, while the other parameters are antisqueezed. This is visualized in Fig.~\ref{Fig2}(b) via the G space and coherence plane. Compared to Fig.~\ref{Fig2}(a), the phase $\arg(G_2+\mathrm{i}G_3)$ has a sharper value, but the magnitude $|G_2+\mathrm{i}G_3|$ shows more uncertainty. 

Next, we turn to examine the implications of the coherence squeezing at the slits on the intensity fringe distributions at screen $\mathcal{B}$. Let still $\alpha$ be real. With the condition $\bar{n}\gg\sinh q$, Eq.~(\ref{B-intensity}) yields the mean intensity
\begin{equation}
     I\approx|K|^2\bar{n}[1+\cos(k\Delta r)].\label{intensity}
\end{equation}
Equation (\ref{intensity}) portrays maximal fringe visibility ($V=1$). However, the photon number fluctuates at every point on $\mathcal{B}$, and the nature of these fluctuations is determined by the variables $q$ and $\theta$. For the coherent state $(q=0)$, the variance of Eq.~(\ref{interference-law}) becomes
\begin{equation}
(\Delta I)^2=2|K|^2 I.\label{coherent}
\end{equation}
It is proportional to the intensity and, hence, position dependent via Eq.~(\ref{intensity}). 
The proportionality also implies that Poissonian photon statistics are expected to occur across the fringe pattern. Yet, when $q\neq0$, the intensity fluctuations differ from those in Eq.~(\ref{coherent}). 

For magnitude squeezing $(\theta=0)$, Eq.~(\ref{interference-law}) leads to
\begin{subequations}
\begin{align}
(\Delta I)_{0}^2\approx(\Delta I)^2\mathrm{e}^{-2q}+2|K|^4\bar{n}\sinh(2q)\sin^2(k\Delta r).\label{amplitude-squeezed}
\end{align}
Here, the subscript $0$ stands for the value of $\theta$. The first term shows squeezing with respect to the variance in Eq.~(\ref{coherent}). The second term, however, adds position-dependent noise that cannot be neglected due to its proportionality to $\bar{n}$. For phase squeezing $(\theta=\pi)$
\begin{align}
(\Delta I)_{\pi}^2\approx(\Delta I)^2\mathrm{e}^{2q}-2|K|^4\bar{n}\sinh(2q)\sin^2(k\Delta r).\label{phase-squeezed}
\end{align}
\end{subequations}
Contrary to Eq.~(\ref{amplitude-squeezed}), the first term is greater than that in Eq.~(\ref{coherent}), and the second term carries a minus sign. 

\begin{figure}
    \centering
    \includegraphics[width=0.93\linewidth]{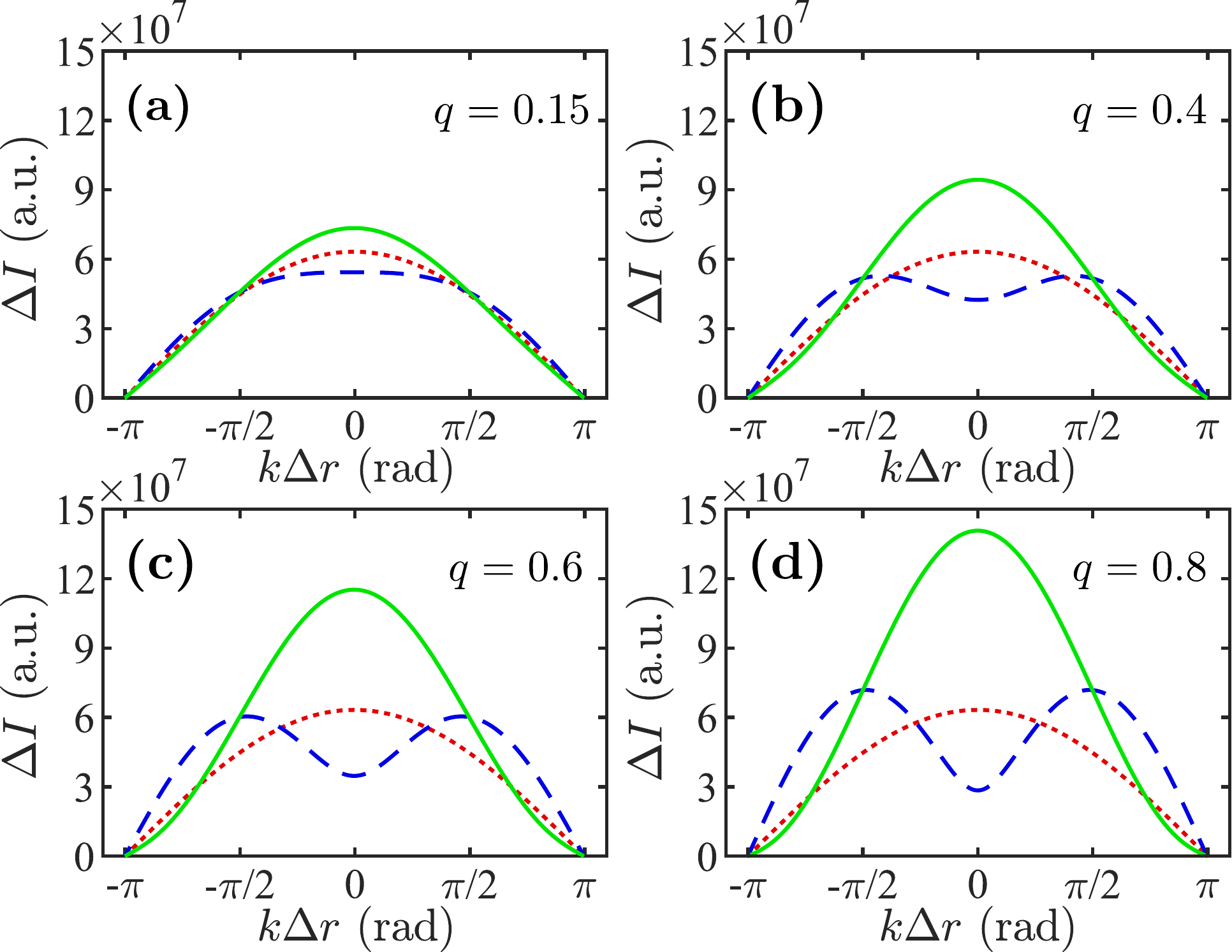}
    \caption{Intensity fluctuations $\Delta I$ from Eqs.~(\ref{coherent})--(\ref{phase-squeezed}) over a fringe-pattern period in terms of the wave number $k$ and path difference $\Delta r$ for selected squeeze parameter values $q$: \textbf{(a)} $0.15$, \textbf{(b)} \textbf{$0.4$}, \textbf{(c) $0.6$}, and \textbf{(d)} $0.8$. The dotted red, blue dashed, and green solid curves correspond to the coherent, amplitude-squeezed, and phase-squeezed states, respectively. We set $K=1$ and $\bar{n}=10^{15}$ in Eqs.~(\ref{coherent})--(\ref{phase-squeezed}).}
    \label{Fig3}
\end{figure}
\begin{figure}
    \centering
    \includegraphics[width=0.85\linewidth]{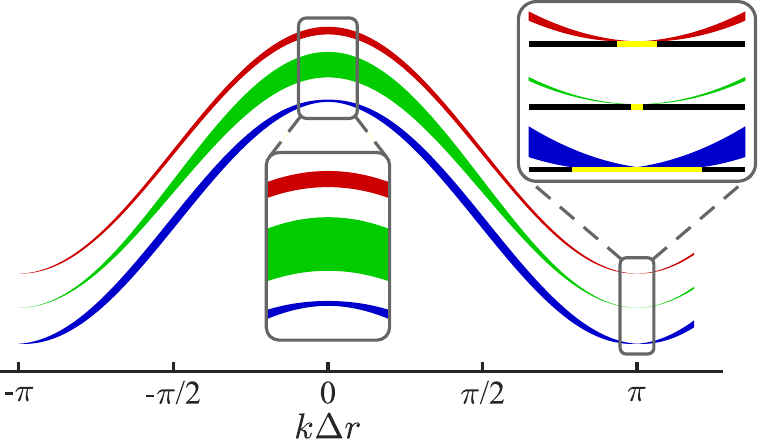}
    \caption{Visualization of the fluctuating intensity fringe 
    patterns based on Eqs.~(\ref{intensity})--(\ref{phase-squeezed}) in terms of the wave number $k$ and path difference $\Delta r$. The red, blue, and green fringes correspond to the coherent, amplitude-squeezed, and phase-squeezed states, respectively. To highlight the differences at the extrema, we have separated the curves vertically and included insets. The yellow lines show the regions where the intensity may reach zero due quantum uncertainty.}
    \label{Fig4}
\end{figure}
Figure~\ref{Fig3} illustrates the intensity fluctuations over a single period of the fringe pattern. The dashed blue curves show the standard deviations acquired from Eq.~(\ref{amplitude-squeezed}) for four different values of $q$. The green curves depict the corresponding values obtained from Eq.~(\ref{phase-squeezed}). Acting as a point of reference, the dotted red curve traces the intensity fluctuations for the coherent state [Eq.~(\ref{coherent})]. We see that amplitude squeezing reduces fluctuations around the intensity maximum ($k\Delta r=0$), but increases them elsewhere. This effect intensifies when the squeeze parameter assumes greater values. In contrast, phase squeezing increases uncertainty around the intensity maximum, but reduces it near the minima $(k\Delta r=\pm\pi)$. 

Figure~\ref{Fig4} visualizes how these intensity fluctuations shape the fringe patterns, with the same coloring as before. Classically, the intensity reaches zero only at $k\Delta r=\pm\pi,\pm3\pi...$, anchoring the interference fringes to exact locations. In the inset, however, we have highlighted in yellow the position ranges on $\mathcal{B}$ where the intensity can vanish due to its quantum fluctuations. As a result, pinpointing the fringe locations is impossible. From Fig.~\ref{Fig4} we observe that the green (blue) fringe has a more well-defined (uncertain) position but a fuzzier (sharper) magnitude than the red reference curve. In this context, the impact of coherence squeezing becomes apparent.
The green fringe represents light that is coherence squeezed at $\mathcal{A}$ in the phase $\arg(G_2+\mathrm{i}G_3)$ and antisqueezed in the magnitude $|G_2+\mathrm{i}G_3|$. As explained below Eq.~(\ref{B-intensity}), the former specifies the locations of the fringes, while the latter determines their magnitude. This type of coherence squeezing thus accounts well for the shape of the green fringe. Similarly, the coherence squeezing of $|G_2+\mathrm{i}G_3|$ and antisqueezing of $\arg(G_2+\mathrm{i}G_3)$ explains the shape of the blue fringe. These connections between the coherence squeezing in the G space and its effects on the interferometric fringe shapes are one of the main findings of our work.

\begin{figure}
    \centering
    \includegraphics[width=0.9\linewidth]{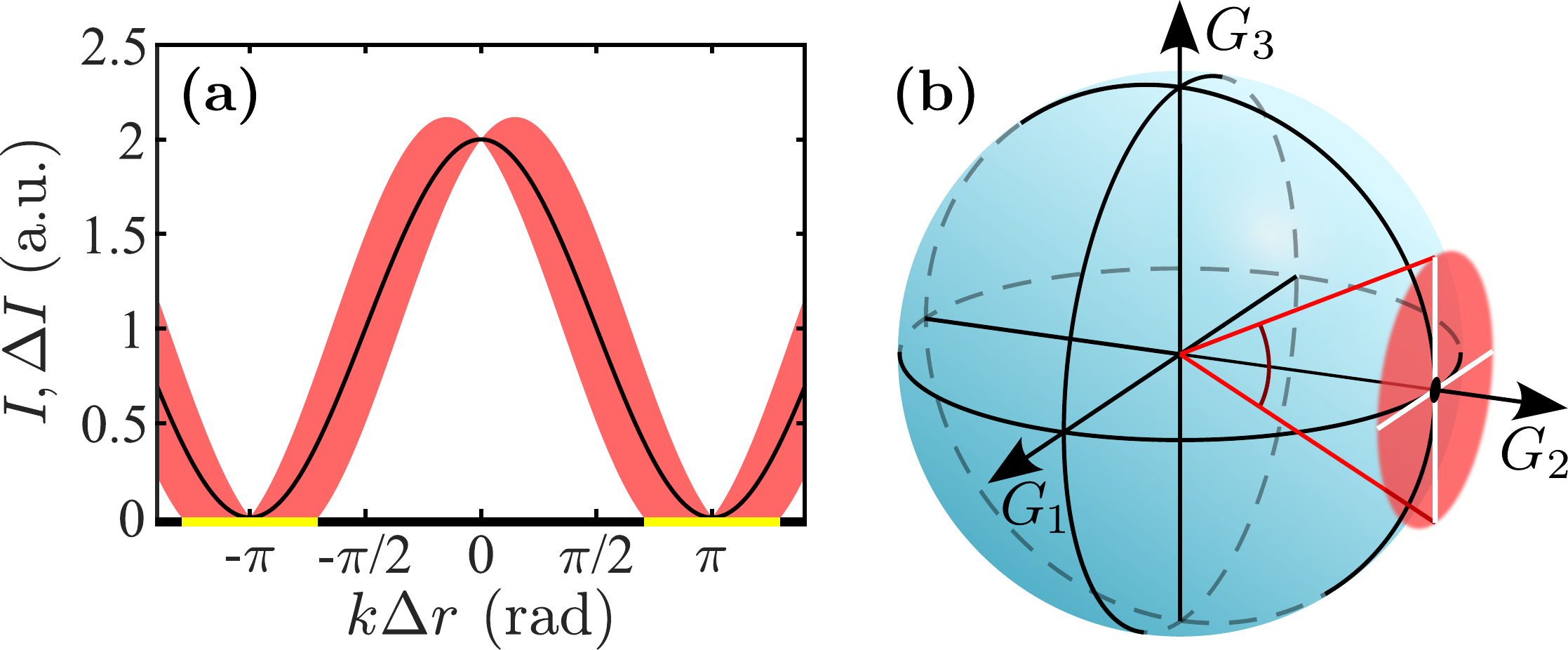}
    \caption{Single-photon interference when $c_1=c_2=1/\sqrt{2}$. \textbf{(a)} Intensity fringe pattern $I\pm \Delta I/2$ obtained from Eq.~(\ref{intensity-single-photon}) for $K=1$ in terms of the wave number $k$ and path difference $\Delta r$. The yellow bars indicate locations at which the uncertainty area touches the horizontal axis. \textbf{(b)} Corresponding G-space representation. The uncertainty in the radial direction ($G_2$) vanishes, yet the uncertainty in $G_1$ and $G_3$ remains. The latter causes the phase $\arg(G_2+\mathrm{i}G_3)$ to exhibit uncertainty, as depicted by the red disc.}
    \label{Fig5}
\end{figure}

\emph{Single-photon light.}---We next consider the pure single-photon state $\ket{\Psi}=c_1\ket{1}_1\!\ket{0}_2+c_2\ket{0}_1\!\ket{1}_2$ in two distinct scenarios. The amplitudes $c_m$ satisfy $|c_1|^2+|c_2|^2=1$, where $|c_m|^2=p_m$ is the probability of the photon traversing slit $m\in\{1,2\}$. In the special case when $c_1=c_2=1/\sqrt{2}$ ($D=0$), Eq.~(\ref{interference-law}) yields $I=|K|^2[1+\cos(k\Delta r)]$. Fluctuations around this mean value are characterized by the standard deviation
\begin{equation}
    \Delta I=|K|^2|\sin(k\Delta r)|.\label{intensity-single-photon}
\end{equation}
Now the fringe pattern displays full visibility $(V=1)$ and the uncertainty $\Delta I$ vanishes at the intensity extrema $k\Delta r=0,\pm\pi,\pm2\pi...$. Figure~\ref{Fig5}(a) illustrates this fringe-pattern behavior, with the yellow bars highlighting the regions where the quantum fluctuations allow the intensity to vanish. Moreover, from Eq.~(\ref{G-operators-monochromatic}) we find $G_0=G_2=1$ and $G_1=G_3=0$, together with the respective fluctuations $\Delta G_0=\Delta G_2=0$ and $\Delta G_1=\Delta G_3=1$. Figure~\ref{Fig5}(b) provides a G-space representation of these uncertainties. We conclude that the uncertainty in the magnitude $|G_2+\mathrm{i}G_3|$ vanishes completely, but the phase $\mathrm{arg}(G_2+\mathrm{i}G_3)$ carries significant uncertainty. Such coherence squeezing and antisqueezing explain the sharp magnitude and blurred positions of the fringes in Fig.~\ref{Fig5}(a).

\begin{figure}
    \centering
    \includegraphics[width=0.9\linewidth]{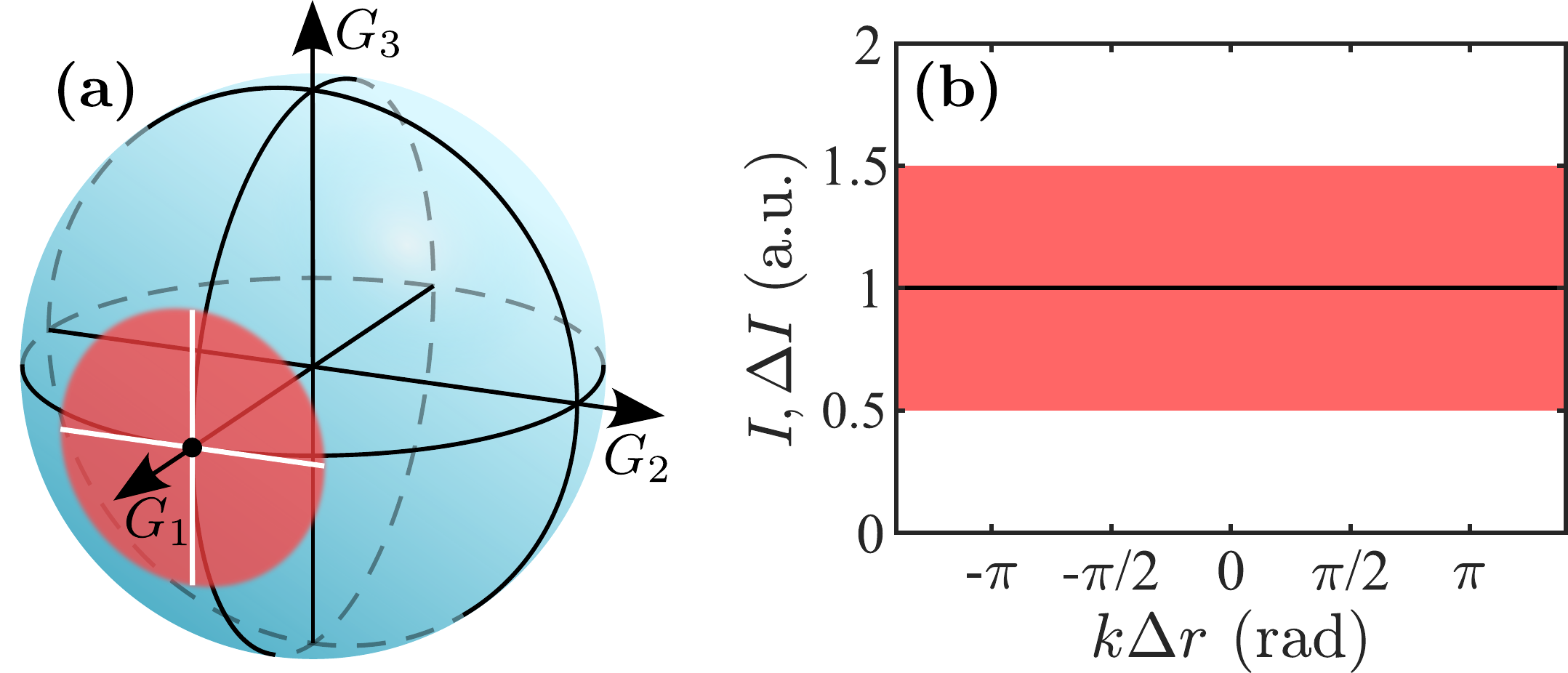}
    \caption{Single-photon case with $p_1=1$ and $p_2=0$. \textbf{(a)} G-space representation: zero uncertainty in $G_1$, but fluctuations in $G_2$ and $G_3$ remain. \textbf{(b)} Constant mean intensity $I$ (black line) surrounded by uniform fluctuations $\Delta I$ (red area) across the observation screen, with $k$ and $\Delta r$ being the wave number and path difference, respectively. We have set $K=1$.}
    \label{Fig6}
\end{figure}

As another example, we analyze the case of $p_1 = 1$ and $p_2 = 0$ $(D = 1)$. Now $G_0=G_1=1$ and $G_2=G_3=0$ as well as $\Delta G_0=\Delta G_1=0$ and $\Delta G_2=\Delta G_3=1$. In other words, the photon traverses slit 1 with full certainty, and, on average, no correlation exists between the one-photon field in slit 1 and the vacuum field in slit 2. However, a uniform distribution of coherence uncertainty exists along the $G_2G_3$ plane, as displayed in Fig.~\ref{Fig6}(a). Hence, the quantity $G_2+\mathrm{i}G_3$ that characterizes the interference fringes has the mean value of zero, but shows uncertainty via its two fluctuating parts. As a consequence, we find from Eq.~(\ref{interference-law}) that the intensity on screen $\mathcal{B}$ is constant and carries uniform fluctuations, $I=\Delta I=|K|^2$, which is illustrated in Fig.~\ref{Fig6}(b).

\emph{Conclusions.}---In summary, we have investigated optical coherence uncertainty in double-slit interference and introduced the concept of coherence squeezing. By constructing Hermitian operators that characterize coherence at the slits, we derived uncertainty relations in terms of distinguishability and visibility. We then formulated a criterion for coherence-squeezed states: Those whose coherence fluctuations fall below the coherent-state level. We showed that both bright squeezed light and single-photon light can meet this criterion, visualized through a Poincaré-sphere-like representation. Interferometric analysis of these cases revealed that coherence squeezing reduces the uncertainty of either the magnitudes or positions of the intensity fringes, with single-photon light exhibiting maximal squeezing in the fringe magnitudes. These results indicate new avenues for quantum-enhanced interferometry by identifying coherence as a fundamental degree of freedom for squeezing alongside phase, amplitude, and polarization.

\emph{Acknowledgments.}---The authors thank Ari T. Friberg for fruitful discussions and correspondence and Elvis Pillinen for assistance with the 3D figures. This work was supported by the Research Council of Finland (Grants No. 354918, No. 349396, and No. PREIN 346518).

\end{document}